\documentstyle[12pt, aaspp4]{article}

\newcommand{\msun}{$M/M_{\odot}\,$}

\begin{document}

\title{Classical Cepheid Pulsation Models. III. The Predictable Scenario.}

\author{Giuseppe Bono\altaffilmark{1}, Vittorio Castellani\altaffilmark{2},
and Marcella Marconi\altaffilmark{3}}

\lefthead{Bono et al.}
\righthead{Cepheid Predictable Scenario}

\altaffiltext{1}{Osservatorio Astronomico di Roma, Via Frascati 33,
00040 Monte Porzio Catone, Italy; bono@coma.mporzio.astro.it}
\altaffiltext{2}{Dipartimento di Fisica, Univ. di Pisa, Piazza Torricelli 2,
56100 Pisa, Italy; vittorio@astr18pi.difi.unipi.it}
\altaffiltext{3}{Osservatorio Astronomico di Capodimonte, Via Moiariello 16,
80131 Napoli, Italy; marcella@na.astro.it}

\normalsize
\vspace*{5.0mm}
\begin{abstract}

Within the current uncertainties in the treatment of the coupling
between pulsation and convection, the limiting amplitude, nonlinear,
convective models appear the only viable approach for providing
theoretical predictions about the intrinsic properties of radial
pulsators. In this paper we present the results of a comprehensive 
set of Cepheid models computed within such theoretical framework 
for selected assumptions on their original chemical composition.

We first discuss the location of the instability strip in the HR
diagram, showing that nonlinear predictions on the effective
temperature of the instability boundaries substantially differ
from similar predictions available in the literature. This 
discrepancy is mainly due to the nonlinear effects introduced by 
the interaction between radial and convective motions and to the 
different physical assumptions adopted for constructing pulsation 
models. We found that both the blue (hot) and the red (cool) 
boundaries of fundamental pulsators appear, for each given 
metallicity, fairly independent of the adopted mass-luminosity 
relation. As a consequence, it turns out that they can be 
approximated over a wide luminosity range by a logarithmic 
relation between stellar luminosity and effective temperature.

We discuss the occurrence of first overtone pulsators and provide
an analytical relation for the effective temperature of the predicted
blue boundary in metal-poor structures. We also found that predicted
fundamental periods based on different assumptions about the
mass-luminosity relation can be all nicely fitted, for each given
metallicity,  by a logarithmic relation which connects the period
of the pulsator to mass, luminosity and effective temperature.

Theoretical predictions concerning pulsation amplitudes in luminosity,
radius, velocity, gravity, and effective temperature are discussed and 
then compared with data available in the current literature. Even 
though these observables are affected by non negligible observational
uncertainties, the agreement between theory and observations is  
rather satisfactory. Finally, we found that the predicted ratio
of the amplitudes in the I and in the V bands appears in very good
agreement with the empirical value ($A_I/A_V=0.6$), with a mild 
dependence on metallicity.

\end{abstract}

\noindent
{\em Subject headings:} galaxies: stellar content -- stars: distances
-- stars: evolution -- stars: oscillations -- stars: variables: Cepheids

\pagebreak
\section{INTRODUCTION}

This is the third paper of a series devoted to the pulsational
properties of classical Cepheids predicted by nonlinear pulsation
models. In paper I (Bono, Marconi  \& Stellingwerf 1999) we presented 
both numerical and physical assumptions adopted for constructing
nonlinear, convective models, and discussed the approach to limit
cycle stability of both fundamental and first overtone pulsators.
By adopting the periods and the modal stability predicted by these
models, paper II (Bono et al. 1999a) was focused on the Period-Luminosity
(PL), the Period-Color (PC), and on the Period-Luminosity-Color (PLC)
relations at selected stellar chemical compositions.
Moreover, Bono, Caputo, \& Marconi (1998) have already discussed the
satisfactory agreement between empirical Cepheid radii and the
corresponding theoretical predictions based on this new theoretical
scenario.

However, nonlinear models do provide many more predictions, such as
the pulsational amplitudes and the morphology of both light and velocity
curves, which can be directly compared with the observed Cepheid
properties. Even though some of these properties were already investigated
on the basis of linear (Chiosi, Wood, \& Capitanio 1993; Simon \& Young 1997;
Saio \& Gautschy 1998; Alibert et al. 1999) and nonlinear
(Carson \& Stothers 1988; Moskalik, Buchler, \& Marom 1992) envelope
models, several empirical evidence such as the change
of luminosity and velocity amplitudes across the instability strip
(Sandage \& Tammann 1971; Pel 1978; Cogan 1980; Fernie 1990) or the
constant amplitude ratio in different photometric bands (Freedman 1988;
Tanvir 1997) have not been enlightened by theoretical insights yet.
In this paper we supply a comprehensive analysis of pulsation predictions
to stimulate further investigations by others working in this field
and to discuss the comparison between these theoretical predictions
and observational data available in the current literature.

In section 2 we discuss the instability boundaries, we compare them with
previous results and discuss the dependence of the boundaries on the
adopted chemical composition as well as on the assumptions about the
mass-luminosity (ML) relation.
In Section 3 we supply suitable analytical relations for fundamental
and first overtone pulsators connecting the period to stellar masses,
luminosities and effective temperatures. Luminosity, radius, gravity,
temperature, and radial velocity amplitudes for fundamental pulsators
are discussed in section 4, while first overtone pulsation amplitudes
are outlined in section 5. In these two sections we also compare
theoretical observables with both photometric and spectroscopic
measurements
available in the literature. Finally, the main findings of this
investigation are  summarized in \S 6, together with some
hints on future developments of the present theoretical framework.

Appendix A gives input parameters and theoretical observables for all 
the computed pulsation models, while Appendix B presents their predicted
light and velocity curves together with a brief discussion on the 
variation of the morphology across the instability strip.

\section{Modal instability and periods}

Limiting amplitude, nonlinear, convective models represent a necessary 
and sufficient condition to assess the modal stability and, in turn, to 
evaluate the edges
of the instability strip for each given stellar structure.
In the case of Cepheid variables, one is dealing with intermediate-mass
stars which cross the instability region at different luminosities
which depend on stellar mass and on chemical composition.
The pulsational investigation thus requires suitable assumptions about 
the ML relation for constraining the predicted luminosity of the model 
for each given value of the stellar mass. For this purpose we adopted 
evolutionary predictions presented by Castellani, Chieffi \& 
Straniero (1992, hereinafter CCS92), which were computed for suitable 
ranges of stellar masses and chemical compositions and under the 
canonical assumption of no convective core overshooting.
Figure 1 shows the comparison between the ML relation adopted in our
computations and the relations adopted in linear investigations by
Chiosi et al. (1993, hereinafter CWC93) and by
Alibert et al. (1999, hereinafter ABHA99). Data plotted in the top
panel for Z=0.02 show that the quoted evolutionary models are in 
excellent agreement. As we will discuss later on, a maximum difference
of the order of $\Delta log L\approx 0.1$ has indeed negligible effects
on the derived pulsational scenario.
The same figure in the  bottom panel shows the dependence of the 
ML relation on the adopted chemical composition. Note that our models 
assume Y=0.28 for Z=0.02 and $\Delta$Y/$\Delta$Z=2.5 for other 
metallicities, in close similarity with the assumptions made by the 
quoted authors. In particular, CWC93 adopted Y=0.25, Z=0.004;
Y=0.25, Z=0.008; and Y=0.30, Z=0.016, while ABHA99 adopted Y=0.25, Z=0.004;
Y=0.25, Z=0.01; and Y=0.28, Z=0.02.  Once again we  find a rather
satisfactory agreement between the different predictions, with a slight
enhanced dependence of ABHA99 relations, to be possibly ascribed to
different input physics adopted in ABHA99 evolutionary models.
However, data plotted in Figure 1 clearly show that we are facing 
a rather well-established evolutionary scenario.

On the basis of the adopted ML relation we explored the pulsation
stability of models at 5, 7, 9, and 11 M$_{\odot}$  with
the effective temperature $T_e$ as a free parameter to be moved by
steps of 100 degrees across the HR diagram. With this procedure
the location of the instability boundaries was evaluated with an 
accuracy of $\pm$50 K. The input physics used for constructing both 
linear and nonlinear envelope models was already described in paper I 
and in paper II as well as in Bono et al. (1998). Here we only mention 
that we adopted OPAL opacities (Iglesias \& Rogers 1996) for temperatures 
greater than $10^4$ K, and Alexander \& Ferguson (1994) molecular 
opacities for lower temperatures.   
In addition an independent set of models was computed for the same 
values of stellar masses but by increasing the predicted luminosities
by $\Delta log L/L_{\odot}=0.25$. According to CWC93, this luminosity shift
allows us to explore an evolutionary scenario which accounts for a
mild convective core overshooting and therefore to investigate the effects
of the mass/luminosity ratio on the pulsation instability. Input parameters
and theoretical results for all models characterized by a stable nonlinear
limit cycle in the fundamental  and/or in the first overtone are
presented in Appendix A, Tables 1-4.

Results concerning the  fundamental edges of the instability strip
are shown in Figure 2 for the two assumed ML relations and
the three different stellar metallicities. Data in this
figure give the plain evidence  that the  instability edges depend 
mainly on luminosity. Noncanonical models -i.e. pulsation models 
constructed by adopting a ML relation based on mild overshooting 
evolutionary models- suggest that the red edges at the highest 
luminosities present a sudden shift toward hotter effective
temperatures, with a sharp decrease in the width of the instability
strip. We find that both canonical and noncanonical boundaries can
be described over a quite large luminosity range by analytical
relations of the type $log T_e \,=\, \alpha + \beta \times log L/L_{\odot}$.
The analytic fits for the red edges were derived by neglecting the
noncanonical point at $ M = 11 M_\odot$, while the blue edge relation 
for Z=0.004 was derived by neglecting the canonical point at 
$M = 5 M_\odot$. Coefficients and errors of both blue and red edge 
relations for the three different chemical compositions are given 
in Table 5.

Moreover, data displayed in Figure 2 strongly support the empirical
evidence originally pointed out by Pel \& Lub (1978) and by
Fernie (1990) that Galactic Cepheids present a "wedge-shaped"
and not a "rectangular-shaped" instability strip. 
A similar behavior for Magellanic Cepheids was originally brought 
out by Martin, Warren, \& Feast (1979) and more recently by Caldwell 
\& Laney (1991). This suggests that the narrowing at the lower 
luminosities is an intrinsic feature of the fundamental instability 
strip which does not depend on metallicity. Figure 2 shows the 
additional evidence that at each given luminosity an increase in the 
metal content shifts the instability strip toward cooler effective 
temperatures. The dependence of PL, PC, and PLC relations on this 
effect was thoroughly described in paper II and therefore it is not 
discussed further.

Figure 3 shows the comparison between our fundamental edges and similar
predictions obtained by CWC93 and by ABHA99 from linear, nonadiabatic,
convective models. We find that at the lowest metallicity our predicted
blue boundaries are located between the two linear ones, with a slope
which appears in good agreement with both CWC93 and ABHA99 results.
However, linear red edges appear much steeper than is predicted by
nonlinear models. Such a discrepancy is not surprising, since the
linear approach does not account for the coupling between pulsation
and convection, and therefore can hardly predict the modal stability
of models located close to the cool edge.
To cope with such a difficulty, linear red edge estimates by CWC93
and by ABHA99 were not fixed at the effective temperature where
the fundamental growth rate attains vanishing values but, more or
less tentatively, at the effective temperature where the growth
rate attains its maximum value (Chiosi et al. 1992; ABHA99).
Present nonlinear results suggest that such an assumption is far
from being adequate, a conclusion further supported by the evidence
that ABHA99 predictions fail to reproduce the already discussed
empirical evidence for a "wedge-shaped" instability strip.
The difference in the blue edges provided by CWC93 and ABHA99 also 
suggests that linear predictions depend on the numerical and 
physical assumptions adopted for constructing pulsation models.

Moreover, one may notice that an increase in the metal content 
causes a flattening in the slope of the nonlinear blue boundary, at
variance with the predicted linear behavior.
Such an occurrence is probably connected with  the evidence that a
metallicity increase shifts the blue edge toward cooler effective
temperatures. As a consequence, metal-rich models show a stronger
dependence on the nonlinear effects introduced by the coupling
between pulsation and convection than metal-poor ones. The reader
interested in a detailed discussion on the dependence of modal
stability and pulsation properties on these effects is referred
to paper I.

Figure 4 shows the location in the HR diagram of first overtone unstable
models we found for the three selected chemical compositions in the lower
portion of the explored luminosity range. In this region for the two
lower metallicities we derived the following linear analytical relations
for first overtone blue boundaries:

$$\log {T_e}^B (Z=0.004)=-0.047(\pm 0.005)\log{L} + 3.954(\pm0.002) \ \ \
\sigma=0.002$$

$$\log {T_e}^B (Z=0.008)=-0.067(\pm 0.004)\log{L} + 4.011(\pm0.001) \ \ \
\sigma=0.001$$

where $L$ is the luminosity (solar units), $T_e$ is the effective
temperature (K), and $\sigma$ is the standard deviation.

\section{Pulsation relations}

During the last few years several authors (CWC93; Simon \& Young 1997;
Saio \& Gautschy 1998) have used  linear, nonadiabatic models to derive
pulsational relation connecting periods to stellar masses and luminosities.
In paper I (see Fig. 53) we already compared theoretical periods
based on both linear and nonlinear models. 
We found that nonlinear periods of high-mass Cepheids are from 1\% to
10\% shorter than the linear ones when moving from the blue to the
red edge of the instability strip. This difference can introduce an
uncertainty  of the order of 0.08 mag in the comparison between
theoretical and empirical PL relations. This finding supports the 
relevance of precise pulsation relations.

Fortunately enough, we find that at each given chemical composition 
the present nonlinear values of canonical and noncanonical fundamental 
periods follow with very good accuracy a linear relation connecting  
$log P$ to the logarithms of the pulsator mass, luminosity, and 
effective temperature.
The coefficients and the errors of these analytical relations
are given in Table 6. Figure 5 shows how accurately these relations 
fit the periods of the computed models. It is now possible to estimate
the effects of uncertainties in the adopted ML relation. 
Going back to Figure 1 we find that at fixed luminosity -the physical
parameter governing the instability boundaries- current canonical  ML
relations predict quite similar mass values. In fact, the difference at
$log L/L_\odot=3.5$ between CCS92 and CWC93 is $\delta log M\approx \pm 0.01$,
while between CCS92 and ABHA99 is $\delta log M\approx \pm 0.02$.
This difference causes a period uncertainty of 2\% and 4\% respectively 
and, in turn, of only few hundreds of magnitude on the distances obtained 
by adopting the theoretical PL relation.
On the other hand, the mass difference between canonical and noncanonical
ML relations is of the order of $\delta log M\approx \pm 0.07$ which
implies a 14\% period uncertainty and, in turn, an uncertainty of
about 0.15 mag on the distance modulus. However, this uncertainty affects
absolute but not relative distance determinations due to the systematic
difference between the two different evolutionary frameworks over the
whole period range.  

Data in Figure 5 shows that for each given effective temperature
an increase in metallicity, in spite of the decrease in luminosity,
causes a small increase in the pulsation period of the order 
of 0.03 dex when moving from Z=0.004 to Z=0.02. However, an increase 
in  metallicity has the major effect of shifting the instability strip
toward lower effective temperatures and therefore longer periods.
As already discussed in Paper II, such a shift plays a key role in
governing the dependence of the PL relation on metallicity.
This shift in temperature also provides a plain explanation of the 
empirical evidence originally
pointed out by Gascoigne (1969, 1974) and more recently by
Sasselov et al. (1997) that Cepheids in the Small Magellanic Cloud
(SMC) are, for a given period, bluer than Cepheids in the LMC.
Alternatively, we predict that at fixed luminosity the period
distribution of SMC Cepheids should be shifted toward shorter
periods when compared with LMC Cepheids. The cumulative period
distributions of Cepheids in LMC and in SMC recently derived
by the EROS project  suggest a similar trend
(Marquette 1998), but we still lack a detailed comparison between
the period distribution of both LMC and SMC Cepheids at fixed 
magnitude bins.

In the explored range of luminosities we found a
relatively small number of first overtone pulsators.
This mode is indeed unstable only in low-mass
($5-7 M_\odot$) and therefore low-luminosity models. Owing to
the mild dependence of first overtone periods on metallicity
we derived a single pulsational relation which
does not account for the metallicity dependence:

$log P = 10.763(\pm0.002) + 0.672(\pm0.002) log L - 3.305(\pm0.039)\log T_e
\
\ \ \sigma=0.002$

where symbols have their usual meaning.


\section{Pulsational amplitudes: fundamental pulsators}

\subsection{Luminosity}

Dating back to the seminal investigation by Sandage \& Tammann (1971,
hereinafter ST71), the luminosity amplitude has been recognized as
a key parameter for assessing the pulsation properties of classical
Cepheids. If amplitudes, as suggested  by the quoted authors (see also
Sandage 1972), depend on the distance from the edges of the instability
strip, this parameter could be used to properly locate a Cepheid within
the strip, and in turn to use a Period-Luminosity-Amplitude relation for
estimating distances. However, empirical evidence concerning the
behavior of the luminosity amplitude inside the strip, the so-called
"amplitude mapping", turned out to be quite controversial.
On the basis of Cepheid samples in four different galaxies, ST71 suggested
that in the period range from $log P \approx 0.40$ to $0.86$ and for
$log P >1.3$ the largest luminosity amplitudes are attained close to the
blue edge, while the trend is reversed for Cepheids with periods ranging
from $log P=0.86$ to $log P=1.3$. A similar conclusion for Galactic 
Cepheids was reached by Cogan (1980), who suggested to move the cut-off 
period from $log P=1.3$ to 1.1.

On the other hand, Madore (1976) suggested, on the basis of empirical
relations provided by Fernie (1970), that classical Cepheids attain the
largest  amplitudes close to the red edge, whereas Butler (1976,1978)
found an opposite trend among MC Cepheids.
More recently Fernie (1990) investigated a large sample of Galactic
Cepheids  suggesting that for a given period
the largest amplitudes are possibly attained close to the center of the
instability strip, even though they are not tightly correlated with the
position in the strip.

Figure 6 shows bolometric amplitudes of both canonical
and noncanonical  models as a function of period for the various given
masses and for the three selected metallicities. In this figure each
sequence of models depicts the predicted  amplitudes across the
instability strip. A glance at the data plotted in Figure 6 discloses 
that for the three explored metallicities the pulsators in the 
short-period range (i.e. with lower masses and luminosities) attain 
their maximum amplitude at the blue edge, whereas toward longer periods 
the amplitudes present a "bell-shaped" distribution.

Such a behavior is  similar to the trend found in fundamental RR Lyrae
pulsators (Bono et al. 1997, hereinafter BCCM97): at the luminosities 
where the fundamental blue boundary falls in a region where the
first overtone is already unstable, the amplitude of fundamental 
pulsators steadly decreases from the blue to the red edge,
whereas at the luminosities in which only the fundamental pulsators
attain a stable nonlinear limit cycle the pulsation amplitudes show
a "bell-shaped" variation from the hot to the cool edge of the instability
strip.  However, we notice that in the large majority of cases the maximum
amplitude is attained only few hundred degrees from the blue boundary, and
then the amplitude steadly decreases from the blue to the red over a large
portion of the strip. Therefore theoretical results appear in reasonable
agreement with the empirical evidence brought out by ST71 and by 
Cogan (1980).

Another feature disclosed by data plotted in Figure 6 is that an increase
in the metal content generally causes a decrease in the maximum amplitudes.
This trend supplies a sound support to the empirical evidence originally 
suggested by Arp \& Kraft (1961) and then confirmed by van Genderen (1978) 
on the basis of a large sample of short-period Galactic and Magellanic 
Cepheids. 
The only exceptions to this rule are the 7 $M_\odot$ (canonical) for Z=0.02 
and the 9 $M_\odot$ (noncanonical) models for Z=0.008 which show a larger 
bolometric amplitude  when compared with more metal-poor models of the 
same mass value.
This peculiarity could be due to a nonlinear behavior of the thermal
properties of the outermost layers caused by the shock propagation
close to the phases of maximum amplitude (see paper I). In fact, as we
will discuss later on, the radial velocity amplitudes of these models
do not show this peculiar behavior.

The sequence of canonical models at 7 $M_\odot$ presents a "double-peaked" 
distribution with two maxima located close to the blue and to the red edge. 
Interestingly enough, these models attain the minimum amplitude at
$log P \approx 1.03$, in remarkable agreement with the minimum at
$log P=1.05\pm0.03$ in the empirical distribution of Fourier parameters
$\phi_{21}$ and $R_{21}$ found in LMC Cepheids by MACHO (Welch et al.
1997) and EROS projects (Beaulieu \& Sasselov 1997).
The physical mechanism(s) which 
govern the appearance of such a phenomenon and the dependence on 
chemical composition will be discussed in a forthcoming paper 
(Bono et al. 1999b).
Finally we note that the period-amplitude variation of this sequence 
supports the local minimum in B amplitudes found in this period range 
by van Genderen (1978) for LMC Cepheids. This finding is very encouraging
since the Bailey diagram (luminosity amplitude vs. period) does not depend 
on distance modulus and on color-temperature relation, and presents a
negligible dependence on bolometric correction scale and on reddening.  

Figure 7 shows the comparison in the Bailey diagram between pulsation 
amplitudes in the V band of Galactic Cepheids collected by Fernie et al. 
(1995) and theoretical predictions for Z=0.008 and Z=0.02. The choice of 
metallicities follows the results of a recent spectroscopic investigation 
by Fry \& Carney (1997), which supports the evidence that Galactic 
calibrating Cepheids cover this metallicity range. Bolometric amplitudes 
were transformed into V amplitudes by adopting the bolometric corrections 
provided by Castelli, Gratton, \& Kurucz (1997a,b). 
We assumed $M_{bol}(\odot)=4.62$.   
Theoretical V amplitudes show the same variation of bolometric amplitudes
across the instability strip and are in satisfactory agreement with
empirical values both at short and at long-periods. The only exception 
are the models at $M = 7 M_\odot$, Z=0.02 and $M = 9 M_\odot$, Z=0.008 
whose V amplitudes are 20-25\% higher than the observed ones.
However, it is noteworthy that the luminosity amplitudes predicted by 
nonlinear {\em radiative} models are at least a factor of two larger 
than those predicted by {\em convective} ones (see Figure 1 in 
Bono \& Marconi 1998).

The period-amplitude behavior of Galactic Cepheids was extensively 
investigated by van Genderen (1974) who found that the envelope line 
of B amplitudes attains a constant value for $0.5 \le log P \le 1.0$ 
and then a rapid increase up to $log P=1.15$. Toward longer periods 
the envelope line attains once again a constant value up to 
$log P=1.5$ and then a steady decrease (see data plotted in his Figure 1).  
This empirical evidence was subsequently confirmed by Laney \& Stobie (1993) 
by adopting infrared J, H, and K amplitudes for 51 Galactic Cepheids.  
Theoretical predictions plotted in Figures 6 and 7 seem to support this 
behavior, but to confirm this evidence further models at solar chemical
composition in the period range $0.7 \le log P \le 1.0$ are needed.

\subsection{Amplitude ratio}

The agreement between theory and observations on the behavior of the
luminosity amplitude inside the instability strip suggested to test
the empirical assumption that the amplitude ratio between the I (Cousins)
and the V (Jonhson) bands is equal to 0.6 mag (Tanvir 1997). This
assumption is generally adopted (Freedman  1988) for deriving the
I-band light curve on the basis of the shape of the V-band light curve
and thus for reducing the number of individual measurements necessary  
for estimating the mean I magnitude.
Amplitude ratios in different photometric bands are also interesting 
because as demonstrated by Coulson \& Caldwell (1989) they can be 
adopted for discovering companions to Cepheids (see also 
Laney \& Stobie 1993).  

Figure 8 shows the predicted $A_I/A_V$ ratio as a function of the 
pulsation period for canonical, fundamental pulsators at the three 
adopted metallicities. Luminosity amplitudes in the I band were 
estimated by adopting the color-temperature relations provided by 
Castelli et al. (1997a,b). Data plotted in this figure show that the 
mean $A_I/A_V$ ratio ranges from $0.64\pm0.03$ at Z=0.004 and Z=0.008
to $0.65\pm0.02$ at Z=0.02.
This result suggests that, within the intrinsic dispersions, the
$A_I/A_V$ ratio is in average constant and slightly higher than the 
empirical value currently adopted.
However, this finding mainly depends on stellar atmosphere models, and 
indeed some numerical experiments performed by artificially enhancing
both the surface gravity and the effective temperature variations along
the pulsation cycle caused an increase in the $A_I/A_V$ ratio smaller 
than 10\%.

\subsection{Radial velocity}

Even though radial velocity amplitudes are available only for a limited
number of Galactic Cepheids, this observable can supply useful information
on the dynamical behavior across the instability strip.
Data listed in Tables 2 and 4 show that the behavior of radial velocity
curves are largely correlated to the already discussed light curves.
The main difference is found in $5 M_\odot$ canonical models at Z=0.004
and Z=0.008, which show very large $\Delta u$ values when compared
with more massive models. Since the bolometric amplitudes do not present
this behavior, it turns out that in these models the bolometric amplitude
over a full pulsation cycle is governed  by temperature variations more 
than by radius variations.

The top panel of Figure 9 shows the comparison between empirical
pulsational velocity amplitudes for Galactic Cepheids provided by
Bersier et al. (1994), Bersier \& Burki (1996) and by
Bersier (1999, private communication) and theoretical predictions
for Z=0.02 and Z=0.008. The bottom
panel of Figure 9 shows a similar comparison but with empirical data
collected by Cogan (1980). In the Cogan sample we identified Cepheid
variables by means of a cross-identification with the database on Galactic
Cepheids provided by Fernie et al. (1995) and among them we selected the
objects presenting accurate velocity amplitudes.
Pulsational velocities plotted in Figure 9 were derived from the
empirical radial velocity amplitudes by adopting a projection factor
of 1.36, as suggested by Bersier \& Burki (1996). However, we notice
that in the literature quite different values were suggested,
with the additional evidence that the projection factor could change
when moving from short to long-period Cepheids (Gieren et al. 1989)
as well as along the pulsation cycle (Sabbey et al. 1995).
We adopted this value since its effect is marginal when compared to 
observational errors affecting non homogeneous spectroscopic measurements
plotted in the bottom panel. In fact, we estimated that the difference
in the velocity amplitude among objects included both in the 
Bersier et al. and in the Cogan samples ranges from few percent 
to more than 20\%.

Figure 9 shows that theoretical predictions appear in reasonable
agreement with observations over the whole period range. In particular,
one may notice that toward longer periods
-$log P=1.5\div1.6$- the velocity amplitudes decrease to approximately
50 km/sec, in agreement with theoretical predictions.
However, from $log P=1.1$ to $log P=1.5$ our computations predict pulsators
with smaller velocity amplitudes that are marginally  supported by current
empirical estimates. We can hardly assess whether this discrepancy is
due to a selection effect or because theoretical amplitudes are too
small close to the instability boundaries. Even by accounting for the
quoted spread in metallicity  the discrepancy between theory and
observations is still present.
In fact, models at Z=0.008 predict both high and small velocity
amplitudes in the period range $log P=1.1\div1.5$.

However, on a very general ground, data plotted in Figure 9 show that
limiting amplitude models which include a proper treatment of the
pulsation/convection interaction seem to solve the long-standing
discrepancy between theoretical and observed velocity amplitudes.
In fact, velocity amplitudes across the instability strip predicted 
by nonlinear radiative models are systematically larger than the 
observed ones (Carson \& Stothers 1988; Moskalik et al. 1992).

\subsection{Radius, gravity and temperature}

Empirical estimates of radius, gravity and temperature amplitudes are
unfortunately rather scanty. In the following we will refer to the last
comprehensive investigations on this subject provided by Pel (1978, 1980)
and based on a large set of photometric data in the Walraven system.
We find that theoretical fractional radius variations  -$\Delta R/R_{ph}$-
for Z=0.02 and Z=0.008 and for periods shorter than 11 d
show -in agreement with empirical data- a decrease when moving from 
the blue to the red edge.
However, predicted  values appear somehow smaller than the observed
ones, since the observed $\Delta R/R_{ph}$ values range from
more than 20\% close to the blue edge to 5\% at the red edge, whereas
theoretical predictions range from 10-15\% close to the blue edge
to 5\% at the red edge.
The origin of such a discrepancy cannot be firmly established since
we lack an estimate of the observational uncertainties as well as
data for long-period Cepheids. Moreover, Pel's results relyed on old
atmosphere models (Kurucz 1975) which could considerably affect 
estimation of these parameters.

Theoretical and observed effective gravity amplitudes -$\Delta g_e$-
appear in reasonable qualitative agreement. The bulk of the empirical
estimates by Pel (1978) for bright Cepheids with $P<11$ d is
around $\Delta g_e\approx$ 0.4, and range roughly from 0.22 to 0.9.
The theoretical ones attain similar values and range from 0.3 to 0.8.
Obviously, spectroscopic investigations can supply tighter
constraints on the accuracy of theoretical predictions.

The agreement we found for gravity amplitudes applies also to fractional
temperature variations. Empirical estimates by Pel (1978) range
approximately from $\Delta \Theta_e$=0.06 to 0.22, while the theoretical
ones from $\Delta \Theta_e$=0.07 to 0.18, where $\Theta_e$ = $5040/T_e$.
This qualitative agreement is further supported by recent spectroscopic
investigations  on the temperature amplitudes in a small sample of Galactic
Cepheids collected by Bersier, Burki, \& Kurucz (1997).
Since the thermal behavior along the pulsation cycle strongly depends on
the coupling between pulsation
and convection, such qualitative agreement in the short-period range
supports, within current observational uncertainties, the treatment
adopted to account for the convective transport.

\section{Pulsational amplitudes: first overtone pulsators}

In Figure 10 from top to bottom are shown bolometric and radial velocity
amplitudes, fractional radius and temperature variations for first
overtone pulsators.
Data plotted in this figure show that in canonical models an increase
in the metal content causes a decrease in the pulsation amplitudes
(see also paper I).
Unfortunately, the small number of unstable first overtone pulsators
does not allow us to find out a clear trend  concerning the change of
the
pulsation amplitudes from the blue to the red edge of the instability
strip. Canonical, metal-poor models at $5 M_\odot$ show a linear
decrease in the bolometric amplitude and in the fractional temperature
variation, while amplitudes of more metal-rich and noncanonical models
present across the instability strip the characteristic "bell-shaped"
variation we already found for first overtone RR Lyrae models (BCCM97).

In order to supply a useful theoretical framework for first overtone
identification among field stars, the light and velocity curves of all
first overtone models are presented in appendix A.
Data plotted in Figures 10-21 show that the shape of both light and
velocity curves of first overtone models are almost sinusoidal and
resemble the empirical
light curves for {\em s}-Cepheids recently observed by the EROS project in
the bar of the LMC (Beaulieu et al. 1995). However, it is worth noting that
the shape of the light curve of canonical models at $M=5 M_\odot$ and
Z=0.004 are not sinusoidal, and indeed the rising branch is steeper than
the decreasing branch. Moreover both canonical and noncanonical models
located in the middle of the instability strip and with period ranging
from P=1.934 d to P=3.811 d also show a well defined bump just before
the luminosity maximum. This feature is also present in models at Z=0.008
with periods ranging from P=2.076 d to P=3.816 d.

Empirical light curve Fourier parameters $\phi_{21}$ of {\em s}-Cepheids
show a sudden jump close to P=3.2 d when plotted as a function of the
pulsation period. This abrupt change was explained as a $2:1$
resonance between the first and the fourth overtone
(Antonello \& Poretti 1986; Petersen 1989). Current radiative
hydrodynamical models fail to reproduce this empirical behavior,
and in particular the appearance around the quoted period of the bump
along the light curve (Kienzle et al. 1999, and references therein).
A detailed analysis of this phenomenon is beyond the scope of this
investigation. However, the evidence that the bump along the light
curve of first overtone pulsators is located in the right period
range suggests that nonlinear, convective models can shed new light
on this long-standing problem.

CORAVEL radial velocity amplitudes for Galactic first overtone
pulsators have been recently collected by Bersier et al. (1994)
and by Kienzle et al. (1999, and references therein).
Theoretical predictions appear once again in reasonable
agreement with empirical data, since close to $log P=0.35$
both data sets attain values of the order of 20-25 km/sec.
The same agreement is found for fractional temperature variations,
and indeed both empirical and theoretical observables for Galactic
Cepheids (Bersier et al. 1997) with period shorter than
$log P=0.5$ range approximately from $\Delta T_e/T_e$=0.05 to 0.09.
However, we do not put forward the comparison between theory and
observations since first overtone models need to be extended to
lower stellar masses before firm conclusions on the behavior of
this mode within the instability strip can be reached.

\section{Summary and conclusions}

In this paper we presented the large set of pulsational properties
predicted by our classical Cepheid, limiting amplitude, nonlinear,
convective models constructed by adopting three different assumptions
on chemical composition. We first discussed the location of the instability
strip in the H-R digram, and showed that the instability edges predicted
by nonlinear models are substantially different from the linear ones.
This difference is mainly caused by the nonlinear effects of the coupling
between pulsation and convection which is not included in linear,
nonadiabatic, convective models.
For each given metallicity we find that both blue and red boundaries 
of fundamental pulsators appear fairly independent of the adopted 
ML relation, and indeed they can be approximated over a quite large 
luminosity range by an analytical relation between $log L$ and $log T_e$.

The occurrence of first overtone pulsators in the lower mass range is
also discussed and the linear analytical relations for the blue boundaries
of metal-poor structures -Z=0.004, Z=0.008- are provided as well.
We also found that predicted fundamental periods, at fixed chemical
composition, can be nicely fitted by an analytical relation connecting
the logarithmic period to $log M$, $log L$ and $log T_e$ independently
of the assumption about the ML relation.
These findings supply straightforward theoretical support to the
use of Cepheid PL relation for estimating distances, since the topology
of the instability strip presents, at fixed chemical composition, a
negligible dependence on the ML relation. A similar conclusion, though 
based on empirical evidence, was reached by Tanvir (1997, and
references therein).

Theoretical predictions concerning luminosity, radius, velocity, gravity,
and temperature amplitudes are discussed in connection with observational
data available in the current literature. As a whole, we found a rather
satisfactory agreement, within the present large observational
uncertainties, between theory and observations. The exhaustive discussion
on theoretical observables presented in this investigation was mainly
aimed at stimulating further detailed comparison with spectroscopic and
photometric data which can supply tight constraints on the adequacy
and consistency of the input physics adopted for constructing
nonlinear pulsation models.

We also mention that Cepheids for which estimates of both stellar 
mass and effective temperature (or color) are available, such as
Cepheids in stellar clusters (Bono \& Marconi 1997) or in binary
systems (B\"ohm-Vitense et al. 1998, and references therein), can
supply useful suggestions on the accuracy of predicted periods,
and in turn on the ML relation which governs these objects.
In fact, together with Cepheids in LMC and SMC stellar clusters 
which are a fundamental laboratory for studying their evolutionary 
and pulsational properties, current empirical estimates suggest 
that the incidence of binaries among field Cepheids ranges from 30\% 
(Evans 1992) to more than 50\% (Szabados \& Pont 1998).

We conclude that the nonlinear theoretical framework appears to be
the only approach which can supply a reliable description of the
pulsation behavior
in radial pulsators. Moreover, it turns out that limiting amplitude
calculations are a fundamental requirement for estimating on a firm
basis both the modal stability and the intrinsic properties to be
compared with observed variable stars (see also Ya'ari \& Tuchman 1999).

We are particularly grateful to D. Bersier for sending us Cepheid radial 
velocity data in electronic form and for new data in advance of publication 
as well as for a detailed reading of an early draft of this paper. 
We are also grateful to D. Alves, F. Caputo, D. Laney and N. Panagia 
for many insightful discussions on Cepheid properties. 
In addition, we acknowledge an anonymous referee for some useful 
remarks that served to improve the paper. This work was supported, 
in part, by the Ministero dell'Universit\`a e della Ricerca Scientifica 
e Tecnologica -Cofinanziamento 98- under the project "Stellar Evolution". 
Partial support by Agenzia Spaziale Italiana is also acknowledged.

\appendix

\section{Predicted observables}

In order to provide a useful theoretical framework which accounts for
the systematic properties of Cepheids to be compared with actual
properties of observed variables, Tables 1-4 summarize the observables
of all limiting amplitude, nonlinear, convective models we computed.

Tables 1 and 2 list both input parameters and pulsational amplitudes
for first overtone and fundamental pulsators constructed by adopting a
ML relation based on evolutionary models which neglect the convective
core overshooting.
In particular, columns 1, 2, and 3 report the stellar mass (solar units),
the logarithmic luminosity (solar units), and the static effective
temperature (K) adopted for each model. From left to right the other
quantities listed in these tables are:
4) nonlinear period (d);
5) mean radius (solar units);
6) fractional radius oscillation i.e.
$\Delta R/R_{ph}\,=\, (R^{max}\,-\,R^{min}) / R_{ph}$ where $R_{ph}$
is the photospheric radius;
7) radial velocity amplitude (km $s^{-1}$) i.e.
$\Delta u \,=\, u^{max}\,-\,u^{min}$;
8) bolometric amplitude (mag.) i.e.
$\Delta M_{bol} \,=\, M_{bol}^{max}\,-\,M_{bol}^{min}$;
9) logarithmic amplitude of static gravity i.e.
$\Delta log g_s \,=\, log g_s^{max}\,-\, log g_s^{min}$;
10) logarithmic amplitude of effective gravity i.e.
$\Delta log g_{eff} \,=\, log g_{eff}^{max}\,-\,log g_{eff}^{min}$
where $g_{eff}  \,=\, GM/R^2\,+\, du/dt$;
11) temperature amplitude (K) i.e.
$\Delta T \,=\, T^{max}\,-\,T^{min}$ where $T$ is the temperature
of the outer boundary;
12) effective temperature amplitude (K) i.e.
$\Delta T_e \,=\, T_e^{max}\,-\,T_e^{min}$ where $T_e$ is
derived from the surface luminosity.
With the exception of the effective gravity, the quantities reported
in column 4 to 11 are referred to the surface zone.
The mean effective gravity of the outermost layers was estimated by
adopting the procedure suggested by Bono, Caputo \& Stellingwerf (1994).
The temperature amplitudes listed in column 10 and 11 have been rounded
to the nearest 50 K.
At fixed stellar mass, luminosity level and chemical composition
the temperature of the blue (red) edges can be estimated by increasing
(decreasing) the effective temperature of the hottest (coolest)
unstable model by 50 K.

Tables 3 and 4 report the same quantities of Tables 1 and 2 but refer
to envelope models constructed by adopting a ML relation typical of
evolutionary models which account for a mild convective core
overshooting (see paper I for details).

\section{Light and velocity curves}

The morphology of both light and velocity curves play a key role for
assessing Cepheid pulsation modes and also for constraining the
pulsation behavior of these variables across the instability strip.
Figures 11-34 show the light (left panel) and velocity (right panel)
variations throughout two consecutive cycles. Solid and dashed lines 
refer to fundamental and first overtone pulsators respectively.
Figures 11-22 show the Cepheid sequences constructed by adopting 
a canonical ML relation, while figures 23-34 the sequences based on 
a noncanonical one. 
However, figures 11-14 as well as 23-26 (\msun=5,7,9,11; Y=0.25; Z=0.004) 
are published in the printed edition, while figures 15-18 and 27-30 
(\msun=5,7,9,11; Y=0.25; Z=0.008), together with figures 19-22 and 31-34 
(\msun=5,7,9,11; Y=0.28; Z=0.02) are only available in the on-line edition.    

The first extensive nonlinear investigations on the shape of theoretical
light and velocity curves of Cepheids date back to Christy (1966, 1975
and references therein), Stobie (1969) and more recently to
Carson \& Stothers (1988) and Moskalik et al. (1992). Light and velocity
curves presented in Figures 11-34, when compared with similar predictions
available in the literature, do not show spurious features such as spikes
and/or ripples throughout the pulsation cycle (see also paper I).
In this context it is worth mentioning that Christy (1975) on the basis
of leading physical arguments, suggested that the main Cepheid features
can be correlated to the {\em Christy parameter} $P/(R/R_\odot)$.
Even though Christy's predictions were based on nonlinear, radiative,
small amplitude models, current convective models support his findings 
concerning the systematics of Cepheid properties.
In fact we found, in agreement with Christy, that the transition from
first overtone to fundamental pulsators takes place close to
$P/(R/R_\odot)\approx0.1$
Moreover, our models show that the appearance of the bump along the
decreasing branch takes place close to $P/(R/R_\odot)\approx0.12$ and
falls at earlier pulsation phases when moving toward longer period
Cepheids. Christy predicted a similar behavior but for slightly larger
$P/(R/R_\odot)$ values.

Unfortunately our models do not reach very long periods ($P > 130$ d)
and therefore we cannot assess on firm basis whether Cepheids
in this region of the instability strip are characterized by an
irregular behavior and by very large amplitudes. Finally, we mention
that our models do not present RV Tauri characteristics, i.e. alternations
of deep and shallow minima in both light and velocity curves in the period
range $25-40$ d, as suggested by Moskalik \& Buchler (1991) and by Moskalik
et al. (1992). Since these calculations were performed by adopting
similar ML relations, we suspect that this difference is caused by
different assumptions on the energy transport mechanism (radiative 
versus convective).

\pagebreak

\pagebreak

\figcaption [] {Comparison between different ML relations. {\em Top
Panel}: ML relations at solar chemical compositions provided by
CWC93 (dashed line), by ABHA99 (long-dashed line) and by CCS92 (filled
squares). These relations are based on evolutionary models which neglect
convective core overshooting, while the open squares mimic the behavior
of a ML relation which include the convective core overshooting.
{\em Bottom panel:} Same as the top panel, but referred to ML relations
derived by adopting different chemical compositions (see labeled values).}

\figcaption [] {Instability strip for fundamental pulsators in the
H-R diagram as a function of metal content. Solid and dashed lines
refer to the blue and the red edge respectively. The edges of models
based on the canonical ML relations are plotted as filled circles, while
models based on the noncanonical ML relation with triangles. The error
bars show the theoretical uncertainty ($\pm 50$ K) on the effective
temperature of the edges.}

\figcaption [] {Comparison in the H-R diagram between linear and
nonlinear (solid lines) fundamental edges at fixed chemical composition
(see labeled values). The edges estimated by CWC93 (dashed lines) and by
ABHA99 (filled circles) are based on linear, nonadiabatic, convective
models. See text for further details.}

\figcaption [] {Distribution of first overtone pulsators (open circles)
in the H-R diagram as a function of the metal content. The solid and the
dashed lines show the edges of fundamental and first overtone pulsators
respectively.}

\figcaption [] {Predicted fundamental periods as a function of the
effective temperature for both canonical (triangles) and noncanonical
(squares) models at different chemical compositions and
stellar masses (see labeled values). The solid lines refer to the
fundamental periods obtained by adopting the analytical relations
connecting the logarithmic period to $log L$, $log M$, and $log T_e$.}

\figcaption [] {Bolometric amplitudes for fundamental pulsators
as a function of period. Solid and dashed lines refer to canonical
and noncanonical models respectively. Models characterized by different
stellar masses are plotted with different symbols. The three panels
refer to models constructed by adopting different chemical compositions
(see labeled values).}

\figcaption [] {Comparison of theoretical amplitudes for Z=0.008 
(solid lines) and Z=0.02 (dashed lines) with empirical amplitudes for 
Galactic Cepheids collected by Fernie et al. (1995) on the $log P-A_V$  
plane. The mass values are specified close to the amplitude sequences.}    

\figcaption [] {Amplitude ratio in the I (Cousins) and in the V (Johnson)
bands for canonical, fundamental models versus period. From top to bottom
the panels refer to models with different chemical compositions
(see labeled values).}

\figcaption [] {Comparison between observed and theoretical radial
velocity amplitudes as a function of period. Empirical radial velocity
amplitudes were transformed into true pulsational amplitudes by adopting
a projection factor of 1.36. Solid and dashed lines refer to sets of
models constructed by adopting Z=0.02 and Z=0.008 respectively and to
different stellar masses (see labeled values). {\em Top Panel}: Velocity
amplitudes for Galactic Cepheids measured by Bersier et al. (1994)
and by Bersier (1999, private communication). {\em Bottom Panel}: 
As the top panel but for velocity amplitudes collected by Cogan (1980).}

\figcaption [] {First overtone pulsation amplitudes as function of
the pulsation period. From top to bottom the panels show the bolometric
and radial velocity amplitude, the fractional radius and temperature
variations. From left to right the panels refer to different chemical
compositions (see labeled values). Models of different stellar masses
are plotted  with different symbols.}

\figcaption [] {Bolometric light curves (left panel) and velocity
(right panel) curves as a function of the pulsation phase. The
curves plotted in this figure refer to a sequence of models
constructed at fixed mass, luminosity and chemical composition
(see labeled values) and canonical ML relation. Dashed and solid 
lines show the luminosity and the velocity variation over two 
pulsation cycles of first overtone and fundamental pulsators 
respectively. In the left panel are also listed the nonlinear 
periods (d), while on the right panel the static effective 
temperatures (K). Positive values along the velocity curves denote 
expansion phases, while negative values denote contraction phases.}

\figcaption [] {Similar to the previous figure, but for \msun=7 models.}

\figcaption [] {Similar to the previous figure, but for \msun=9 models.} 

\figcaption [] {Similar to the previous figure, but for \msun=11 models.} 

\figcaption [] {As in Fig. 11, but for more metal-rich models 
(Z=0.008 against Z=0.004).}
 
\figcaption [] {Similar to the previous figure, but for \msun=7 models.}

\figcaption [] {Similar to the previous figure, but for \msun=9 models.} 

\figcaption [] {Similar to the previous figure, but for \msun=1 models.} 

\figcaption [] {As in Fig. 15, but for models constructed by adopting  
higher helium (Y=0.28 against Y=0.25) and metal (Z=0.02 against Z=0.008) 
contents.}
 
\figcaption [] {Similar to the previous figure, but for \msun=7 models.}

\figcaption [] {Similar to the previous figure, but for \msun=9 models.} 

\figcaption [] {Similar to the previous figure, but for \msun=11 models.} 

\figcaption [] {As in Fig. 11, but for models constructed by adopting
a noncanonical ML relation.}

\figcaption [] {Similar to the previous figure, but for \msun=7 models.}  

\figcaption [] {Similar to the previous figure, but for \msun=9 models.}

\figcaption [] {Similar to the previous figure, but for \msun=11 models.}

\figcaption [] {As in Fig. 15, but for models constructed by adopting
a noncanonical ML relation.}

\figcaption [] {Similar to the previous figure, but for \msun=7 models.}  

\figcaption [] {Similar to the previous figure, but for \msun=9 models.}

\figcaption [] {Similar to the previous figure, but for \msun=11 models.}

\figcaption [] {As in Fig. 19, but for models constructed by adopting
a noncanonical ML relation.}

\figcaption [] {Similar to the previous figure, but for \msun=7 models.}  

\figcaption [] {Similar to the previous figure, but for \msun=9 models.}

\figcaption [] {Similar to the previous figure, but for \msun=11 models.}

\end{document}



\tablewidth{0pt} 
\begin{deluxetable}{ccccccccccccc}
\tablecaption{First overtone pulsators (canonical ML relation).}
\label{tbl-4} 
\tablehead{
\colhead{$M$\tablenotemark{a}}&
\colhead{$log L$\tablenotemark{b}}&
\colhead{$T_e$\tablenotemark{c}}&
\colhead{$P$\tablenotemark{d}}& 
\colhead{$\log{\overline{R}}$\tablenotemark{e}}& 
\colhead{$\Delta R/R_{ph}$\tablenotemark{f}}& 
\colhead{$\Delta u$\tablenotemark{g}}&
\colhead{$\Delta M_{bol}$\tablenotemark{h}}& 
\colhead{$\Delta {\log{g_s}}$\tablenotemark{i}}&  
\colhead{$\Delta {\log{g_e}}$\tablenotemark{j}}&
\colhead{$\Delta T$\tablenotemark{k}}&
\colhead{$\Delta T_{e}$\tablenotemark{l}}&
\colhead{$KE$\tablenotemark{m}} \nl    
\colhead{(1)}&
\colhead{(2)}&
\colhead{(3)}&
\colhead{(4)}&
\colhead{(5)}&
\colhead{(6)}&
\colhead{(7)}&
\colhead{(8)}& 
\colhead{(9)}& 
\colhead{(10)}& 
\colhead{(11)}&
\colhead{(12)}&
\colhead{(13)}}    
 \startdata
     \multicolumn{13}{c}{Y=0.25 Z=0.004} \nl 
   &      &      &        & &      &       &      &     &     &     &     \nl
5.0& 3.07 & 6400 & 1.7732 &1.450 &0.066 & 51.55 &0.934 &0.06 &0.73 & 1200& 1500& 2.418(40)\nl
5.0& 3.07 & 6200 & 1.9338 &1.477 &0.075 & 60.62 &0.782 &0.06 &0.79 & 950 & 1150& 4.458(40)\nl
5.0& 3.07 & 6000 & 2.1879 &1.505 &0.064 & 49.66 &0.492 &0.06 &0.64 & 600 & 750 & 1.960(40)\nl
5.0& 3.07 & 5900 & 2.3107 &1.518 &0.053 & 39.53 &0.383 &0.05 &0.50 & 500 & 600 & 8.535(39) \nl 
7.0& 3.65 & 6000 & 5.3420 &1.794 &0.026 & 15.50 &0.247 &0.02 &0.21 & 300 & 400 & 5.640(39)\nl
7.0& 3.65 & 5900 & 5.6686 &1.810 &0.067 & 38.20 &0.621 &0.06 &0.55 & 750 & 950 & 3.606(40)\nl
7.0& 3.65 & 5800 & 6.0112 &1.826 &0.085 & 48.72 &0.678 &0.07 &0.71 & 800 & 1000& 8.991(40)\nl
7.0& 3.65 & 5700 & 6.3668 &1.842 &0.088 & 49.86 &0.562 &0.08 &0.72 & 700 & 850 & 1.175(41)\nl
7.0& 3.65 & 5600 & 6.7294 &1.855 &0.078 & 42.12 &0.440 &0.07 &0.62 & 550 & 700 & 8.649(40)\nl 
   &      &      &        & &      &       &      &     &     &     &     \nl
     \multicolumn{13}{c}{Y=0.25 Z=0.008} \nl 
   &      &      &        &      &       &      &     &     &     &     \nl
5.0& 3.07 & 6370 & 1.8014 &1.453 &0.048 & 37.59 &0.657 &0.04 &0.49 & 850 & 1050& 1.268(40)\nl
5.0& 3.07 & 6300 & 1.8682 &1.463 &0.063 & 49.58 &0.815 &0.05 &0.65 & 1050& 1300& 2.346(40)\nl
5.0& 3.07 & 6100 & 2.0758 &1.491 &0.066 & 52.32 &0.599 &0.06 &0.66 & 700 & 850 & 2.280(40)\nl
5.0& 3.07 & 6000 & 2.1905 &1.505 &0.060 & 46.70 &0.485 &0.05 &0.59 & 600 & 750 & 1.445(40)\nl
5.0& 3.07 & 5900 & 2.3139 &1.519 &0.051 & 51.42 &0.394 &0.04 &0.48 & 500 & 600 & 7.305(39)\nl
7.0& 3.65 & 5800 & 6.0167 &1.825 &0.049 & 27.70 &0.408 &0.04 &0.38 & 500 & 600 & 2.692(40)\nl
7.0& 3.65 & 5700 & 6.3789 &1.840 &0.063 & 34.46 &0.459 &0.05 &0.49 & 550 & 700 & 4.815(40)\nl 
   &      &      &        &      &       &      &     &     &     &     \nl
     \multicolumn{13}{c}{Y=0.28 Z=0.02} \nl 
   &      &      &        &      &       &      &     &     &     &     \nl
5.0& 3.07 & 6100 & 2.0731 &1.490 &0.027 & 20.28 &0.276 &0.02 &0.24 & 350 & 400 & 2.011(39) \nl
5.0& 3.07 & 6000 & 2.1847 &1.504 &0.037 & 27.80 &0.344 &0.03 &0.33 & 400 & 500 & 3.335(39)\nl
5.0& 3.07 & 5900 & 2.3122 &1.519 &0.030 & 21.58 &0.242 &0.03 &0.26 & 300 & 350 & 1.569(39)\nl 
\enddata
\tablenotetext{a}{ Stellar mass (solar units).  
\hspace*{0.5mm} $^b$ Logarithmic luminosity (solar units).  
\hspace*{0.5mm} $^c$ Effective temperature (K).
\hspace*{0.5mm} $^d$ Period (day). 
\hspace*{0.5mm} $^d$ Logarithmic mean radius (solar units). 
\hspace*{0.5mm} $^f$ Fractional radius variation.  
\hspace*{0.5mm} $^g$ Radial velocity amplitude (Km$s^{-1}$).
\hspace*{0.5mm} $^h$ Bolometric amplitude (mag).
\hspace*{0.5mm} $^i$ Amplitude of logarithmic static gravity.  
\hspace*{0.5mm} $^j$ Amplitude of logarithmic effective gravity.
\hspace*{0.5mm} $^k$ Surface temperature variation (K).   
\hspace*{0.5mm} $^l$ Effective temperature variation (K).
\hspace*{0.5mm} $^m$ Total kinetic energy (erg).}
\end{deluxetable}

\clearpage 
\tablewidth{0pt} 
\begin{deluxetable}{ccccccccccccc}
\tablecaption{Fundamental Pulsators (Canonical ML relation).}
\label{tbl-5} 
\tablehead{
\colhead{$M$\tablenotemark{a}}&
\colhead{$log L$\tablenotemark{b}}&
\colhead{$T_e$\tablenotemark{c}}&
\colhead{$P$\tablenotemark{d}}& 
\colhead{$\log{\overline{R}}$\tablenotemark{e}}& 
\colhead{$\Delta R/R_{ph}$\tablenotemark{f}}& 
\colhead{$\Delta u$\tablenotemark{g}}&
\colhead{$\Delta M_{bol}$\tablenotemark{h}}& 
\colhead{$\Delta {\log{g_s}}$\tablenotemark{i}}&  
\colhead{$\Delta {\log{g_e}}$\tablenotemark{j}}&
\colhead{$\Delta T$\tablenotemark{k}}&
\colhead{$\Delta T_{e}$\tablenotemark{l}}&
\colhead{$KE$\tablenotemark{m}}\nl   
\colhead{(1)}&
\colhead{(2)}&
\colhead{(3)}&
\colhead{(4)}&
\colhead{(5)}&
\colhead{(6)}&
\colhead{(7)}&
\colhead{(8)}& 
\colhead{(9)}& 
\colhead{(10)}& 
\colhead{(11)}&
\colhead{(12)}&
\colhead{(13)} }    
\startdata
     \multicolumn{13}{c}{Y=0.25  Z=0.004} \nl 
   &      &      &  &      &      &       &      &     &     &     &     \nl
5.0& 3.07 & 5900 & 3.1920 &1.523 &0.152 & 82.34 &0.810 &0.13 &0.91 & 1050& 1300& 3.734(41)\nl
5.0& 3.07 & 5700 & 3.5812 &1.549 &0.119 & 60.12 &0.563 &0.10 &0.67 & 800 & 950 & 1.806(41)\nl
5.0& 3.07 & 5600 & 3.8008 &1.561 &0.067 & 32.78 &0.265 &0.06 &0.33 & 400 & 500 & 1.935(40)\nl 
7.0& 3.65 & 5900 & 7.9616 &1.811 &0.099 & 42.44 &0.820 &0.09 &0.52 & 1000& 1200& 2.549(41)\nl
7.0& 3.65 & 5800 & 8.4499 &1.826 &0.111 & 49.91 &0.829 &0.10 &0.56 & 950 & 1200& 3.099(41)\nl
7.0& 3.65 & 5700 & 8.9766 &1.841 &0.110 & 50.23 &0.723 &0.10 &0.59 & 800 & 950 & 3.732(41)\nl
7.0& 3.65 & 5600 & 9.5358 &1.855 &0.100 & 45.59 &0.599 &0.09 &0.56 & 650 & 800 & 2.941(41)\nl
7.0& 3.65 & 5500 & 10.1497&1.869 &0.093 & 41.48 &0.460 &0.08 &0.49 & 550 & 650 & 3.815(41)\nl
7.0& 3.65 & 5400 & 10.7996&1.883 &0.072 & 29.72 &0.302 &0.06 &0.37 & 400 & 500 & 2.326(41)\nl
7.0& 3.65 & 5350 & 11.1330&1.890 &0.062 & 24.38 &0.245 &0.05 &0.32 & 300 &400  & 1.843(41)\nl
7.0& 3.65 & 5300 & 11.5081&1.896 &0.036 & 12.55 &0.137 &0.03 &0.18 & 200 & 250 & 4.106(40)\nl
7.0& 3.65 & 5250 & 11.8828&1.904 &0.035 & 13.22 &0.141 &0.03 &0.17 & 200 & 200 & 1.189(40)\nl
7.0& 3.65 & 5200 & 12.2781&1.912 &0.051 & 20.43 &0.215 &0.05 &0.23 & 300 & 250 & 1.120(40)\nl
9.0& 4.00 & 5800 & 14.3252&1.999 &0.019 & 6.92  &0.129 &0.02 &0.08 & 150 & 200 & 8.904(39)\nl
9.0& 4.00 & 5700 & 15.2461&2.014 &0.066 & 22.20 &0.390 &0.06 &0.34 & 400 & 500 & 8.362(40)\nl
9.0& 4.00 & 5600 & 16.2950&2.031 &0.104 & 39.72 &0.671 &0.09 &0.50 & 800 & 950 & 1.881(41)\nl
9.0& 4.00 & 5500 & 17.4438&2.049 &0.151 & 58.44 & 1.017&0.13 &0.78 & 1200& 1450& 6.149(41)\nl
9.0& 4.00 & 5400 & 18.7160&2.069 &0.191 & 71.31 & 1.117&0.16 & 1.06& 1350& 1650& 1.258(42)\nl
9.0& 4.00 & 5300 & 20.0209&2.087 &0.223 & 80.35 & 1.056&0.19 & 1.16& 1300& 1600& 2.168(42)\nl
9.0& 4.00 & 5200 & 21.3641&2.103 &0.233 & 80.88 & 1.167&0.20 & 1.15& 1400& 1750& 2.996(42)\nl
9.0& 4.00 & 5100 & 22.7370&2.115 &0.216 & 70.62 & 1.030&0.19 &0.95 & 1250& 1550& 1.405(42)\nl
9.0& 4.00 & 5000 & 24.1723&2.127 &0.185 & 56.39 &0.802 &0.16 &0.77 & 1000& 1250& 7.006(41)\nl
9.0& 4.00 & 4900 & 25.7072&2.137 &0.134 & 38.07 &0.486 &0.12 &0.51 & 650 & 800 & 2.893(41)\nl
11.0& 4.40& 5500 & 34.7262&2.249 &0.131 & 32.20 &0.641 &0.23 &0.53 & 800 & 1000& 1.485(42)\nl
11.0& 4.40& 5400 & 37.4530&2.270 &0.216 & 51.31 & 1.056&0.30 &0.93 & 1250& 1550& 3.658(42)\nl
11.0& 4.40& 5300 & 40.1659&2.288 &0.246 & 58.89 & 1.201&0.34 & 1.03& 1350& 1650& 4.436(42)\nl
11.0& 4.40& 5200 & 42.9870&2.304 &0.254 & 60.88 & 1.163&0.37 & 1.00& 1200& 1500& 4.562(42)\nl
11.0& 4.40& 5100 & 45.9495&2.319 &0.250 & 59.61 & 1.040&0.40 &0.95 & 1050& 1300& 4.541(42)\nl
11.0& 4.40& 5000 & 49.0565&2.333 &0.235 & 55.46 &0.879 &0.43 &0.88 & 900 & 1100& 4.297(42)\nl
11.0& 4.40& 4900 & 52.1530&2.345 &0.216 & 49.37 &0.745 &0.45 &0.79 & 850 & 1000& 3.453(42)\nl
11.0& 4.40& 4800 & 55.8722&2.359 &0.194 & 43.16 &0.660 &0.47 &0.69 & 750 & 900 & 2.673(42)\nl
11.0& 4.40& 4700 & 59.6443&2.379 &0.171 & 37.04 &0.565 &0.49 &0.59 & 600 & 750 & 1.919(42)\nl
11.0& 4.40& 4600 & 63.6795&2.385 &0.145 & 30.40 &0.453 &0.51 &0.48 & 500 & 650 & 1.226(42)\nl
11.0& 4.40& 4500 & 67.9329&2.397 &0.115 & 23.30 &0.327 &0.52 &0.36 & 400 & 450 & 6.507(41)\nl
11.0& 4.40& 4400 & 72.3487&2.409 &0.075 & 14.47 &0.183 &0.53 &0.22 & 250 & 300 & 2.539(41)\nl 
   &      &      &        & &      &       &      &     &     &     &     \nl
     \multicolumn{13}{c}{Y=0.25  Z=0.008} \nl 
   &      &      &        &   &   &       &      &     &     &     &     \nl
5.0& 3.07 & 5900 & 3.2417 &1.523 & 0.136 & 73.80 &0.875 &0.12 &0.80 & 1050& 1300& 2.982(42)\nl
5.0& 3.07 & 5800 & 3.4292 &1.536 & 0.130 & 68.01 &0.693 &0.11 &0.75 & 950 & 1150& 2.991(41)\nl
5.0& 3.07 & 5700 & 3.6315 &1.550 & 0.117 & 58.94 &0.596 &0.10 &0.64 & 800 & 1000& 1.440(41)\nl
5.0& 3.07 & 5600 & 3.8487 &1.563 & 0.095 & 46.10 &0.442 &0.08 &0.48 & 600 & 750 & 4.500(40)\nl
5.0& 3.07 & 5500 & 4.0813 &1.576 & 0.061 & 28.81 &0.248 &0.05 &0.29 & 350 & 450 & 1.126(40)\nl 
7.0& 3.65 & 5800 & 8.5734 &1.824 & 0.066 & 30.71 &0.513 &0.06 &0.34 & 650 & 850 & 8.266(40)\nl
7.0& 3.65 & 5700 & 9.1025 &1.840 & 0.089 & 38.84 &0.650 &0.08 &0.45 & 750 & 900 & 2.591(40)\nl
7.0& 3.65 & 5600 & 9.6741 &1.856 & 0.094 & 40.83 &0.595 &0.08 &0.48 & 700 & 850 & 2.433(41)\nl
7.0& 3.65 & 5500 & 10.2660&1.869 & 0.087 & 36.43 &0.464 &0.08 &0.45 & 550 & 650 & 2.691(41)\nl
7.0& 3.65 & 5450 & 10.6009&1.877 & 0.080 & 35.15 &0.381 &0.07 &0.42 & 500 & 550 & 2.395(41)\nl
7.0& 3.65 & 5400 & 10.9473&1.884 &0.062 & 23.18 &0.296 &0.05 &0.33 & 350 & 400 & 1.227(41)\nl
7.0& 3.65 & 5350 & 11.2904&1.892 &0.058 & 20.75 &0.282 &0.05 &0.31 &300  & 350 & 8.270(40)\nl
7.0& 3.65 & 5300 & 11.6768&1.899 &0.064 & 25.89 &0.326 &0.06 &0.34 & 350 & 450 & 3.036(40)\nl
7.0& 3.65 & 5200 & 12.5072&1.918 &0.113 & 47.61 &0.601 &0.10 &0.57 & 700 & 850 & 8.099(40)\nl
7.0& 3.65 & 5100 & 13.3591&1.935 &0.140 & 55.15 &0.723 &0.12 &0.69 & 850 & 1050& 1.432(41)\nl
7.0& 3.65 & 5000 & 14.2372&1.949 &0.131 & 46.10 &0.562 &0.12 &0.58 & 700 & 850 & 1.553(41)\nl 
9.0& 4.00 & 5600 & 16.5176&2.030 &0.080 & 28.06 &0.507 &0.07 &0.33 & 600 & 750 & 2.063(41)\nl
9.0& 4.00 & 5500 & 17.6816&2.050 &0.147 & 52.86 &0.980 &0.13 &0.68 & 1200& 1500& 7.027(41)\nl
9.0& 4.00 & 5400 & 18.9332&2.069 &0.191 & 66.91 & 1.236&0.16 &0.96 & 1500& 1900& 1.360(42)\nl
9.0& 4.00 & 5300 & 20.2593&2.088 &0.221 & 75.24 & 1.340&0.19 & 1.11& 1600& 1950& 2.018(42)\nl
9.0& 4.00 & 5200 & 21.6398&2.104 &0.231 & 77.42 & 1.171&0.20 & 1.02& 1350& 1650& 1.637(42)\nl
9.0& 4.00 & 5100 & 23.0796&2.118 &0.222 & 70.52 & 1.021&0.19 &0.95 & 1250& 1550& 1.125(42)\nl
9.0& 4.00 & 5000 & 24.6201&2.132 &0.204 & 60.43 &0.880 &0.18 &0.86 & 1100& 1350& 6.842(41)\nl
9.0& 4.00 & 4900 & 26.2864&2.145 &0.181 & 49.98 &0.714 &0.16 &0.74 & 900 & 1100& 4.545(41)\nl
9.0& 4.00 & 4800 & 28.1229&2.160 &0.158 & 39.90 &0.539 &0.14 &0.63 & 700 & 900 & 3.145(41)\nl
9.0& 4.00 & 4700 & 30.0358&2.173 &0.130 & 31.64 &0.354 &0.12 &0.47 & 500 & 650 & 3.282(41)\nl
11.0& 4.40& 5200 & 43.4044&2.299 &0.177 & 40.32 &0.805 &0.34 &0.66 & 900 & 1150& 1.888(42)\nl
11.0& 4.40& 5100 & 46.5814&2.318 &0.212 & 47.21 &0.932 &0.39 &0.79 & 1000& 1250& 2.617(42)\nl
11.0& 4.40& 5000 & 49.8826&2.327 &0.221 & 49.22 &0.927 &0.43 &0.78 & 950 & 1150& 3.072(42)\nl
11.0& 4.40& 4900 & 53.4119&2.350 &0.218 & 48.63 &0.853 &0.46 &0.76 & 850 & 1050& 2.991(42)\nl
11.0& 4.40& 4850 & 55.2678&2.358 &0.217 & 48.51 &0.827 &0.47 &0.76 &800  & 1000& 3.064(42)\nl
11.0& 4.40& 4800 & 57.1738&2.352 &0.206 & 45.34 &0.750 &0.48 &0.71 & 750 &900  & 2.645(42)\nl
11.0& 4.40& 4750 & 59.1773&2.373 &0.204 & 44.98 &0.736 &0.49 &0.71 & 750 & 900 & 2.631(42)\nl
11.0& 4.40& 4700 & 60.9618&2.379 &0.196 & 42.49 &0.678 &0.51 &0.67 & 700 & 850 & 2.372(42)\nl
11.0& 4.40& 4600 & 65.6808&2.394 &0.180 & 38.51 &0.594 &0.54 &0.60 & 600 & 750 & 2.004(42)\nl
11.0& 4.40& 4500 & 70.4054&2.409 &0.161 & 34.11 &0.519 &0.56 &0.53 & 550 & 650 & 1.622(42)\nl
11.0& 4.40& 4400 & 75.4589&2.423 &0.139 & 28.81 &0.436 &0.58 &0.44 & 450 & 550 & 1.178(42)\nl
11.0& 4.40& 4300 & 80.6113&2.436 &0.112 & 22.47 &0.332 &0.60 &0.34 & 350 & 450 & 7.290(41)\nl
11.0& 4.40& 4200 & 85.7446&2.447 &0.079 & 15.10 &0.207 &0.61 &0.23 & 250 & 300 & 3.011(41)\nl
   &      &      &        &  &    &       &      &     &     &     &     \nl
     \multicolumn{13}{c}{Y=0.28  Z=0.02} \nl 
   &      &      &        &   &   &       &      &     &     &     &     \nl
5.0& 3.07 & 5900 & 3.3169 &1.520 &0.085 & 43.74 &0.639 &0.07 &0.43 & 800 & 950 & 4.841(40)\nl
5.0& 3.07 & 5800 & 3.5140 &1.535 & 0.095 & 48.08 &0.621 &0.08 &0.49 & 750 & 900 & 5.496(40) \nl
5.0& 3.07 & 5700 & 3.7190 &1.549 & 0.096 & 47.62 &0.549 &0.08 &0.49 & 650 & 800 & 4.998(40)\nl
5.0& 3.07 & 5600 & 3.9453 &1.564 & 0.089 & 42.48 &0.429 &0.08 &0.44 & 550 & 700 & 3.377(40)\nl
5.0& 3.07 & 5500 & 4.1863 &1.579 & 0.077 & 35.57 &0.344 &0.07 &0.37 & 450 & 550 & 1.902(40)\nl 
7.0& 3.65 & 5400 & 11.3422&1.889 & 0.131 & 51.12 &0.805 &0.11 &0.55 & 950 & 1200& 1.342(41)\nl
7.0& 3.65 & 5300 & 12.1039&1.908 & 0.168 & 65.67 & 1.039&0.14 &0.76 & 1200& 1500& 2.445(41)\nl
7.0& 3.65 & 5200 & 12.9389&1.927 & 0.194 & 74.01 & 1.077&0.17 &0.91 & 1250& 1550& 3.582(41)\nl
7.0& 3.65 & 5100 & 13.8115&1.945 & 0.206 & 75.13 & 1.042&0.18 &0.98 & 1250& 1550& 4.371(41)\nl
7.0& 3.65 & 5000 & 14.7615&1.961 & 0.208 & 71.84 &0.989 &0.18 &0.94 & 1200& 1450& 6.131(41)\nl
7.0& 3.65 & 4900 & 15.7542&1.976 & 0.196 & 65.34 &0.868 &0.17 &0.82 & 1000& 1250& 5.824(41)\nl
7.0& 3.65 & 4800 & 16.8212&1.988 & 0.155 & 48.38 &0.576 &0.14 &0.58 & 700 & 850 & 2.660(41)\nl
7.0& 3.65 & 4700 & 17.9605&2.000 & 0.042 & 12.08 &0.118 &0.04 &0.14 & 150 & 200 & 1.282(40)\nl 
9.0& 4.00 & 5100 & 23.7088&2.114 & 0.141 & 38.80 &0.614 &0.12 &0.51 & 750 & 900 & 3.653(41)\nl
9.0& 4.00 & 5000 & 25.4006&2.132 & 0.171 & 45.22 &0.680 &0.15 &0.65 & 850 & 1050& 4.112(41)\nl
9.0& 4.00 & 4900 & 27.2110&2.150 & 0.182 & 45.79 &0.643 &0.16 &0.69 & 800 & 950 & 4.406(41)\nl
9.0& 4.00 & 4700 & 31.3104&2.185 & 0.192 & 45.35 &0.599 &0.16 &0.69 & 700 & 850 & 6.551(41)\nl
9.0& 4.00 & 4500 & 36.0629&2.217 & 0.172 & 39.48 &0.496 &0.18 &0.58 & 550 & 700 & 6.327(41)\nl
9.0& 4.00 & 4400 & 38.7582&2.230 & 0.124 & 27.87 &0.317 &0.19 &0.40 & 400 & 450 & 3.497(41)\nl 
11.0& 4.40& 4800 & 59.5490&2.363 & 0.070 & 13.73 &0.229 &0.44 &0.19 & 250 & 300 & 1.544(41)\nl
11.0& 4.40& 4600 & 69.0000&2.402 & 0.177 & 34.61 &0.633 &0.55 &0.54 & 600 & 750 & 1.344(42)\nl
11.0& 4.40& 4400 & 79.8194&2.435 & 0.177 & 35.14 &0.615 &0.62 &0.54 & 550 & 650 & 1.689(42)\nl
11.0& 4.40& 4300 & 86.0767&2.451 & 0.162 & 31.49 &0.538 &0.64 &0.48 & 450 & 600 & 1.538(42)\nl
11.0& 4.40& 4200 & 92.6401&2.466 & 0.142 & 27.10 &0.452 &0.67 &0.43 & 400 & 500 & 1.310(42)\nl
11.0& 4.40& 4100 & 99.3362&2.480 & 0.117 & 21.64 &0.346 &0.69 &0.35 & 300 & 400 & 1.182(42)\nl
11.0& 4.40& 4000 &106.1928&2.493 & 0.086 & 15.49 &0.247 &0.70 &0.25 & 250 & 300 & 1.115(42)\nl
11.0& 4.40& 3900 &112.9544&2.503 & 0.048 & 8.43  &0.130 &0.71 &0.13 & 150 & 150 & 5.137(41)\nl
\enddata
\tablenotetext{a}{ Stellar mass (solar units).  
\hspace*{0.5mm} $^b$ Logarithmic luminosity (solar units).  
\hspace*{0.5mm} $^c$ Effective temperature (K).
\hspace*{0.5mm} $^d$ Period (day). 
\hspace*{0.5mm} $^e$ Logarithmic mean radius (solar units). 
\hspace*{0.5mm} $^f$ Fractional radius variation.  
\hspace*{0.5mm} $^g$ Radial velocity amplitude (Km$s^{-1}$).
\hspace*{0.5mm} $^h$ Bolometric amplitude (mag).
\hspace*{0.5mm} $^i$ Amplitude of logarithmic static gravity.  
\hspace*{0.5mm} $^j$ Amplitude of logarithmic effective gravity.
\hspace*{0.5mm} $^k$ Surface temperature variation (K).   
\hspace*{0.5mm} $^l$ Effective temperature variation (K).
\hspace*{0.5mm} $^m$ Total kinetic energy (erg).}
\end{deluxetable}

\clearpage 
\tablewidth{0pt} 
\begin{deluxetable}{ccccccccccccc}
\tablecaption{First overtone pulsators (noncanonical ML relation).}
\label{tbl-4} 
\tablehead{
\colhead{$M$\tablenotemark{a}}&
\colhead{$log L$\tablenotemark{b}}&
\colhead{$T_e$\tablenotemark{c}}&
\colhead{$P$\tablenotemark{d}}& 
\colhead{$\log{\overline{R}}$\tablenotemark{e}}& 
\colhead{$\Delta R/R_{ph}$\tablenotemark{f}}& 
\colhead{$\Delta u$\tablenotemark{g}}&
\colhead{$\Delta M_{bol}$\tablenotemark{h}}& 
\colhead{$\Delta {\log{g_s}}$\tablenotemark{i}}&  
\colhead{$\Delta {\log{g_e}}$\tablenotemark{j}}&
\colhead{$\Delta T$\tablenotemark{k}}&
\colhead{$\Delta T_{e}$\tablenotemark{l}}&
\colhead{$KE$\tablenotemark{m}} \nl    
\colhead{(1)}&
\colhead{(2)}&
\colhead{(3)}&
\colhead{(4)}&
\colhead{(5)}&
\colhead{(6)}&
\colhead{(7)}&
\colhead{(8)}& 
\colhead{(9)}& 
\colhead{(10)}& 
\colhead{(11)}&
\colhead{(12)}&
\colhead{(13)} }    
 \startdata
     \multicolumn{13}{c}{Y=0.25 Z=0.004} \nl 
   &      &      &  &      &      &       &      &     &     &     &     \nl
5.0& 3.30 & 6200 & 3.0308 &1.592 &0.040 & 26.92 &0.412 &0.03 &0.35 & 500 & 650 & 4.633(39)\nl
5.0& 3.30 & 6100 & 3.2039 &1.606 &0.053 & 35.63 &0.461 &0.05 &0.50 & 550 & 700 & 7.077(39)\nl
5.0& 3.30 & 5800 & 3.8107 &1.650 &0.072 & 46.48 &0.459 &0.06 &0.60 & 550 & 700 & 1.333(40)\nl
5.0& 3.30 & 5600 & 4.2466 &1.677 &0.056 & 32.86 &0.300 &0.05 &0.44 & 400 & 500 & 1.073(40)\nl
   &      &      &        &  &    &       &      &     &     &     &     \nl
     \multicolumn{13}{c}{Y=0.25 Z=0.008} \nl 
   &      &      &        &   &   &       &      &     &     &     &     \nl
5.0& 3.30 & 6100 & 3.2094 &1.605 &0.034 & 22.07 &0.326 &0.03 &0.30 & 400 & 500 & 1.996(39)\nl
5.0& 3.30 & 6000 & 3.3994 &1.620 &0.046 & 28.72 &0.409 &0.04 &0.41 & 500 & 600 & 3.695(39)\nl 
5.0& 3.30 & 5800 & 3.8160 &1.651 &0.076 & 47.74 &0.555 &0.07 &0.61 & 650 & 850 & 1.612(40)\nl
5.0& 3.30 & 5600 & 4.2726 &1.678 &0.056 & 32.01 &0.325 &0.05 &0.43 & 400 & 500 & 9.831(39)\nl
\enddata
\tablenotetext{a}{ Stellar mass (solar units).  
\hspace*{0.5mm} $^b$ Logarithmic luminosity (solar units).  
\hspace*{0.5mm} $^c$ Effective temperature (K).
\hspace*{0.5mm} $^d$ Period (day). 
\hspace*{0.5mm} $^e$ Logarithmic mean radius (solar units). 
\hspace*{0.5mm} $^f$ Fractional radius variation.  
\hspace*{0.5mm} $^g$ Radial velocity amplitude (Km$s^{-1}$).
\hspace*{0.5mm} $^h$ Bolometric amplitude (mag).
\hspace*{0.5mm} $^i$ Amplitude of logarithmic static gravity.  
\hspace*{0.5mm} $^j$ Amplitude of logarithmic effective gravity.
\hspace*{0.5mm} $^k$ Surface temperature variation (K).   
\hspace*{0.5mm} $^l$ Effective temperature variation (K).
\hspace*{0.5mm} $^m$ Total kinetic energy (erg).}
\end{deluxetable}

\clearpage 
\tablewidth{0pt} 
\begin{deluxetable}{ccccccccccccc}
\tablecaption{Fundamental pulsators (noncanonical ML relation).}
\label{tbl-5} 
\tablehead{
\colhead{$M$\tablenotemark{a}}&
\colhead{$log L$\tablenotemark{b}}&
\colhead{$T_e$\tablenotemark{c}}&
\colhead{$P$\tablenotemark{d}}& 
\colhead{$\log{\overline{R}}$\tablenotemark{e}}& 
\colhead{$\Delta R/R_{ph}$\tablenotemark{f}}& 
\colhead{$\Delta u$\tablenotemark{g}}&
\colhead{$\Delta M_{bol}$\tablenotemark{h}}& 
\colhead{$\Delta {\log{g_s}}$\tablenotemark{i}}&  
\colhead{$\Delta {\log{g_e}}$\tablenotemark{j}}&
\colhead{$\Delta T$\tablenotemark{k}}&
\colhead{$\Delta T_{e}$\tablenotemark{l}}&
\colhead{$KE$\tablenotemark{m}}\nl   
\colhead{(1)}&
\colhead{(2)}&
\colhead{(3)}&
\colhead{(4)}&
\colhead{(5)}&
\colhead{(6)}&
\colhead{(7)}&
\colhead{(8)}& 
\colhead{(9)}& 
\colhead{(10)}& 
\colhead{(11)}&
\colhead{(12)}&
\colhead{(13)} }    
\startdata
     \multicolumn{13}{c}{Y=0.25  Z=0.004} \nl 
   &      &      &  &      &      &       &      &     &     &     &     \nl
5.0& 3.30 & 6100 & 4.4638 &1.607 &0.112 & 50.97 & 1.044&0.10 &0.67 & 1300& 1600& 2.022(41)\nl
5.0& 3.30 & 6000 & 4.7249 &1.623 & 0.127 & 60.38 & 1.044&0.11 &0.72 & 1250& 1550& 2.362(41)\nl
5.0& 3.30 & 5900 & 5.0102 &1.637 & 0.130 & 62.13 &0.936 &0.11 &0.72 & 1050& 1300& 1.975(41)\nl
5.0& 3.30 & 5700 & 5.6156 &1.665 &0.115 & 54.03 &0.589 &0.10 &0.63 & 700 & 850 & 1.064(41) \nl
5.0& 3.30 & 5500 & 6.3499 &1.692 &0.077 & 33.32 &0.348 &0.07 &0.39 & 450 & 550 & 8.231(40)\nl
5.0& 3.30 & 5400 & 6.7478 &1.705 &0.045 & 19.46 &0.176 &0.04 &0.20 & 250 & 300 & 1.324(39)\nl
7.0& 3.85 & 5900 & 11.8936&1.910 &0.008 & 2.78  &0.050 &0.01 &0.04 & 50  & 100 & 3.051(39)\nl
7.0& 3.85 & 5800 & 12.6386&1.925 &0.060 & 19.82 &0.380 &0.05 &0.30 & 400 & 500 & 2.583(40)\nl
7.0& 3.85 & 5700 & 13.4916&1.941 &0.109 & 39.33 &0.739 &0.09 &0.54 & 850 & 1050& 1.083(41)\nl
7.0& 3.85 & 5600 & 14.4477&1.959 &0.155 & 57.12 & 1.070&0.13 &0.80 & 1300& 1600& 3.565(41)\nl
7.0& 3.85 & 5500 & 15.4727&1.978 &0.196 & 70.16 & 1.239&0.17 & 1.08& 1550& 1900& 7.756(41)\nl
7.0& 3.85 & 5400 & 16.5601&1.996 &0.228 & 79.75 & 1.222&0.20 & 1.22& 1450& 1750& 1.314(42)\nl
7.0& 3.85 & 5300 & 17.6617&2.012 &0.235 & 80.13 & 1.120&0.20 & 1.12& 1400& 1700& 1.447(42)\nl
7.0& 3.85 & 5200 & 18.7844&2.025 &0.228 & 68.70 &0.972 &0.21 & 1.06& 1300& 1600& 1.737(42)\nl
7.0& 3.85 & 5100 & 19.9887&2.036 &0.198 & 58.45 &0.860 &0.17 &0.86 & 1100& 1350& 4.405(41)\nl
7.0& 3.85 & 5000 & 21.2514&2.048 &0.162 & 44.70 &0.636 &0.15 &0.68 & 850 & 1050& 2.302(41)\nl
7.0& 3.85 & 4900 & 22.6022&2.059 &0.110 & 27.75 &0.339 &0.10 &0.42 & 500 & 600 & 9.621(40)\nl
9.0& 4.25 & 5700 & 25.7807&2.140 &0.004 & 1.12  &0.024 &0.04 &0.21 & 50  & 50  & 4.523(38) \nl
9.0& 4.25 & 5600 & 27.4353&2.156 &0.085 & 21.81 &0.431 &0.07 &0.34 & 550 & 650 & 3.921(41)\nl
9.0& 4.25 & 5500 & 29.6773&2.178 &0.204 & 48.30 & 1.012&0.18 &0.91 & 1250& 1550& 1.956(42)\nl
9.0& 4.25 & 5400 & 31.8640&2.197 &0.249 & 59.07 & 1.231&0.22 & 1.09& 1450& 1800& 2.809(42)\nl
9.0& 4.25 & 5300 & 34.0954&2.214 &0.264 & 63.69 & 1.262&0.23 & 1.08& 1400& 1700& 3.034(42)\nl
9.0& 4.25 & 5200 & 36.4212&2.230 &0.263 & 63.74 & 1.152&0.23 & 1.02& 1200& 1500& 2.930(42)\nl
9.0& 4.25 & 5100 & 38.8574&2.244 &0.253 & 60.83 & 1.013&0.25 &0.98 & 1000& 1200& 3.005(42) \nl
9.0& 4.25 & 5050 & 40.1676&2.250 &0.245 & 58.31 &0.930 &0.26 &0.94 & 1000& 1200& 2.833(42)\nl
9.0& 4.25 & 5000 & 41.4628&2.256 &0.164 & 35.00 &0.540 &0.27 &0.89 & 950 & 1150& 2.611(42)\nl
9.0& 4.25 & 4950 & 42.8265&2.263 &0.224 & 51.29 &0.791 &0.28 &0.84 & 900 & 1100& 2.286(42)\nl
9.0& 4.25 & 4900 & 44.0638&2.268 &0.210 & 47.01 &0.737 &0.29 &0.76 & 850 & 1000& 1.944(42)\nl
9.0& 4.25 & 4800 & 47.1567&2.282 &0.184 & 40.04 &0.628 &0.31 &0.65 & 700 & 900 & 1.377(42)\nl
9.0& 4.25 & 4700 & 50.2887&2.295 &0.164 & 35.00 &0.540 &0.33 &0.56 & 600 & 750 & 1.046(42)\nl
9.0& 4.25 & 4600 & 53.6414&2.307 &0.134 & 27.61 &0.405 &0.35 &0.44 & 500 & 600 & 6.043(41)\nl
9.0& 4.25 & 4500 & 57.1679&2.319 &0.096 & 18.97 &0.253 &0.36 &0.29 & 300 & 400 & 2.660(41)\nl
11.0& 4.65& 5500 & 60.4344&2.371 &0.005 & 1.02  &0.027 &0.43 &0.02 & 50  & 50  & 9.056(39) \nl
11.0& 4.65& 5400 & 63.7238&2.386 &0.025 & 5.29  &0.115 &0.47 &0.08 & 100 & 150 & 7.320(40) \nl
11.0& 4.65& 5300 & 68.5561&2.405 &0.119 & 23.75 &0.503 &0.54 &0.38 & 600 & 700 & 1.208(42)\nl
11.0& 4.65& 5200 & 73.6106&2.422 &0.150 & 29.89 &0.662 &0.58 &0.49 & 800 & 900 & 2.014(42)\nl
11.0& 4.65& 5100 & 79.0768&2.438 &0.162 & 32.19 &0.715 &0.61 &0.56 & 850 & 950 & 2.650(42)\nl
11.0& 4.65& 5000 & 84.9395&2.454 &0.161 & 32.45 &0.702 &0.64 &0.55 & 700 & 850 & 2.824(42)\nl
11.0& 4.65& 4900 & 91.5153&2.465 &0.128 & 24.62 &0.477 &0.66 &0.43 & 500 & 650 & 2.111(42)\nl
11.0& 4.65& 4800 & 98.3615&2.480 &0.112 & 20.69 &0.394 &0.69 &0.39 & 400 & 500 & 3.332(42)\nl
11.0& 4.65& 4700 &105.7185&2.493 &0.088 & 16.39 &0.303 &0.71 &0.34 & 350 & 400 & 2.851(42)\nl 
   &      &      &        &   &   &       &      &     &     &     &     \nl
     \multicolumn{13}{c}{Y=0.25  Z=0.008} \nl 
   &      &      &        & &     &       &      &     &     &     &     \nl
5.0& 3.30 & 6000 & 4.7927 &1.621 &0.099 & 44.11 &0.863 &0.09 &0.52 & 1050& 1300& 9.477(40)\nl
5.0& 3.30 & 5800 & 5.3695 &1.651 &0.114 & 53.75 &0.801 &0.10 &0.59 & 900 & 1100& 1.113(41)\nl
5.0& 3.30 & 5600 & 6.0578 &1.680 &0.098 & 44.37 &0.534 &0.09 &0.50 & 600 & 750 & 1.150(41)\nl
5.0& 3.30 & 5400 & 6.8315 &1.708 &0.070 & 31.40 &0.318 &0.06 &0.35 & 400 & 500 & 4.960(40)\nl
5.0& 3.30 & 5300 & 7.2680 &1.722 &0.045 & 18.30 &0.186 &0.04 &0.21 & 250 & 300 & 1.454(40)\nl 
7.0& 3.85 & 5700 & 13.6624&1.940 &0.053 & 17.38 &0.330 &0.05 &0.21 & 400 & 500 & 5.632(40)\nl
7.0& 3.85 & 5600 & 14.6120&1.959 &0.137 & 47.55 &0.918 &0.11 &0.62 & 1150& 1400& 3.535(41)\nl
7.0& 3.85 & 5500 & 15.6416&1.978 &0.186 & 63.19 & 1.247&0.16 &0.93 & 1550& 1900& 7.375(41)\nl
7.0& 3.85 & 5400 & 16.7248&1.996 &0.218 & 72.31 & 1.377&0.19 & 1.10& 1650& 2050& 1.093(42)\nl
7.0& 3.85 & 5300 & 17.8802&2.013 &0.232 & 75.29 & 1.220&0.20 & 1.03& 1450& 1800& 8.339(41)\nl
7.0& 3.85 & 5200 & 19.0636&2.028 &0.231 & 71.04 & 1.024&0.20 & 1.00& 1250& 1550& 6.341(41)\nl
7.0& 3.85 & 5100 & 20.3106&2.041 &0.213 & 61.46 &0.902 &0.19 &0.91 & 1150& 1400& 4.341(41)\nl
7.0& 3.85 & 5000 & 21.6381&2.055 &0.199 & 52.97 &0.778 &0.17 &0.84 & 1000& 1250& 3.138(41)\nl
7.0& 3.85 & 4900 & 23.1643&2.070 &0.187 & 46.63 &0.645 &0.16 &0.76 & 850 & 1050& 2.893(41)\nl
7.0& 3.85 & 4800 & 24.7424&2.084 &0.172 & 41.49 &0.507 &0.15 &0.65 & 700 & 850 & 3.214(41)\nl
7.0& 3.85 & 4700 & 26.4054&2.096 &0.123 & 29.12 &0.318 &0.11 &0.42 & 450 & 550 & 1.816(41)\nl 
9.0& 4.25 & 5300 & 34.3272&2.208 &0.185 & 46.63 &0.889 &0.17 &0.72 & 1000& 1250& 1.388(42)\nl
9.0& 4.25 & 5200 & 36.8545&2.228 &0.223 & 50.56 & 1.036&0.21 &0.86 & 1150& 1400& 1.847(42)\nl
9.0& 4.25 & 5100 & 39.4494&2.244 &0.234 & 52.87 & 1.030&0.25 &0.86 & 1100& 1350& 1.969(42)\nl
9.0& 4.25 & 5000 & 42.1071&2.260 &0.233 & 52.98 &0.972 &0.28 &0.85 & 950 & 1200& 2.054(42)\nl
9.0& 4.25 & 4900 & 45.1198&2.274 &0.223 & 49.96 &0.831 &0.31 &0.80 & 850 & 1050& 2.011(42)\nl
9.0& 4.25 & 4800 & 48.2842&2.289 &0.188 & 46.00 &0.664 &0.33 &0.74 & 750 & 950 & 1.736(42)\nl
9.0& 4.25 & 4700 & 51.6869&2.304 &0.194 & 42.01 &0.653 &0.36 &0.68 & 700 & 850 & 1.487(42)\nl
9.0& 4.25 & 4600 & 55.3677&2.318 &0.177 & 37.66 &0.585 &0.38 &0.60 & 600 & 750 & 1.211(42)\nl
9.0& 4.25 & 4500 & 59.2946&2.332 &0.156 & 32.64 &0.503 &0.40 &0.52 & 500 & 650 & 9.138(41)\nl
9.0& 4.25 & 4400 & 63.4315&2.346 &0.132 & 26.76 &0.406 &0.42 &0.42 & 400 & 500 & 6.104(41)\nl
9.0& 4.25 & 4300 & 67.6799&2.358 &0.101 & 19.76 &0.288 &0.44 &0.31 & 300 & 400 & 3.128(41)\nl
9.0& 4.25 & 4200 & 71.6850&2.368 &0.006 & 10.56 &0.135 &0.44 &0.16 & 150 & 200 & 8.980(40)\nl
11.0& 4.65& 5200 & 75.0710&2.418 &0.022 & 4.35  &0.094 & 0.53&0.07 & 100 & 150 & 5.107(40)\nl
11.0& 4.65& 5100 & 80.4266&2.435 &0.084 & 15.03 &0.319 &0.59 &0.23 & 400 & 450 & 2.205(41)\nl
11.0& 4.65& 5000 & 86.3803&2.453 &0.113 & 20.53 &0.446 &0.63 &0.34 & 500  &600 & 5.499(41)\nl
11.0& 4.65& 4900 & 92.6797&2.469 &0.125 & 22.85 &0.498 &0.67 &0.38 &500  &650  & 9.420(41)\nl
11.0& 4.65& 4800 &100.2516&2.484 &0.107 & 18.72 &0.370 &0.69 &0.32 & 400 & 500 & 1.257(42)\nl
11.0& 4.65& 4650 &112.1748&2.507 &0.099 & 16.43 &0.330 &0.74 &0.32 & 350 & 400 & 2.226(42)\nl
11.0& 4.65& 4600 &116.7736&2.515 &0.092 & 14.75 &0.315 &0.76 &0.30 & 300 & 350 & 1.676(42)\nl
   &      &      &        &   &   &       &      &     &     &     &     \nl
     \multicolumn{13}{c}{Y=0.28  Z=0.02} \nl 
   &      &      &  &      &      &       &      &     &     &     &     \nl
5.0& 3.30 & 5700 & 5.8631 &1.664 &0.054 &  22.65 &0.364 &0.05 &0.23 & 450 & 550 & 4.435(39)\nl
5.0& 3.30 & 5600 & 6.2267 &1.679 &0.065 & 27.93 &0.414 &0.06 &0.30 & 500 & 600 & 9.396(39)\nl
5.0& 3.30 & 5500 & 6.6094 &1.695 &0.066 & 27.66 &0.369 &0.06 &0.32 & 450 & 550 & 3.774(40)\nl
5.0& 3.30 & 5400 & 7.0148 &1.709 &0.055 & 21.65 &0.258 &0.05 &0.27 & 300 & 400 & 3.340(40)\nl
7.0& 3.85 & 5200 & 19.5354&2.024 &0.162 & 43.76 &0.749 &0.14 &0.61 & 900 & 1150& 2.467(41)\nl
7.0& 3.85 & 5100 & 20.9006&2.042 &0.187 & 48.62 &0.771 &0.16 &0.72 & 950 & 1150& 2.705(41)\nl
7.0& 3.85 & 5000 & 22.3729&2.059 &0.200 & 49.96 &0.735 &0.17 &0.77 & 900 & 1100& 2.968(41)\nl
7.0& 3.85 & 4900 & 23.9614&2.076 &0.206 & 49.68 &0.694 &0.18 &0.78 & 850 & 1050& 3.432(41)\nl
7.0& 3.85 & 4800 & 25.5723&2.093 &0.208 & 48.64 &0.675 &0.18 &0.77 & 800 & 1000& 3.539(41)\nl
7.0& 3.85 & 4700 & 27.5304&2.109 &0.203 & 46.59 &0.632 &0.18 &0.72 & 750 & 950 & 3.950(41)\nl
7.0& 3.85 & 4600 & 29.5150&2.125 &0.191 & 43.20 &0.583 &0.17 &0.66 & 700 & 850 & 3.981(41)\nl
7.0& 3.85 & 4500 & 31.6742&2.140 &0.167 & 37.07 &0.482 &0.15 &0.57 & 550 & 700 & 3.633(41)\nl
7.0& 3.85 & 4400 & 33.9948&2.153 &0.121 & 26.16 &0.304 &0.11 &0.38 & 350 & 450 & 1.903(41)\nl
9.0& 4.25 & 4900 & 46.6106&2.269 &0.036 & 7.24  &0.122 &0.23 &0.07 & 100 & 100 & 2.464(40)\nl
9.0& 4.25 & 4800 & 50.2708&2.292 &0.173 & 34.56 &0.662 &0.33 &0.54 & 650 & 850 & 7.656(41)\nl
9.0& 4.25 & 4700 & 53.9918&2.310 &0.188 & 37.68 &0.695 &0.37 &0.60 & 700 & 850 & 9.653(41)\nl
9.0& 4.25 & 4600 & 58.0595&2.327 &0.191 & 38.20 &0.688 &0.40 &0.61 & 650 & 800 & 1.072(42)\nl
9.0& 4.25 & 4500 & 62.3935&2.343 &0.187 & 37.50 &0.655 &0.44 &0.59 & 600 & 700 & 1.115(42)\nl
9.0& 4.25 & 4300 & 72.2404&2.375 &0.161 & 31.47 &0.533 &0.49 &0.49 & 500 & 600 & 9.806(41)\nl
9.0& 4.25 & 4100 & 83.2813&2.404 &0.111 & 20.26 &0.319 &0.53 &0.33 & 300 & 400 & 5.051(41)\nl
9.0& 4.25 & 3900 & 94.3210&2.425 &0.035 & 5.82  &0.082 &0.55 &0.09 & 100 & 100 & 7.618(40)\nl
11.0& 4.65& 4800 &105.6869&2.488 &0.050 & 7.60  &0.154 &0.68 &0.13 & 150 &200  & 3.203(40)\nl
11.0& 4.65& 4700 &113.7407&2.505 &0.055 &8.19  &0.151 &0.71 &0.13 & 150 & 200 & 2.815(40)\nl
11.0& 4.65& 4600 &122.4423&2.522 &0.077 & 11.30 &0.212 &0.76 &0.19 & 250 & 300 & 3.918(41)\nl
11.0& 4.65& 4500 &132.4339&2.539 &0.082 & 11.55 &0.253 &0.80 &0.23 & 250 & 300 & 7.658(41)\nl
\enddata
\tablenotetext{a}{ Stellar mass (solar units).  
\hspace*{0.5mm} $^b$ Logarithmic luminosity (solar units).  
\hspace*{0.5mm} $^c$ Effective temperature (K).
\hspace*{0.5mm} $^d$ Period (day). 
\hspace*{0.5mm} $^e$ Logarithmic mean radius (solar units).  
\hspace*{0.5mm} $^f$ Fractional radius variation.  
\hspace*{0.5mm} $^g$ Radial velocity amplitude (Km$s^{-1}$).
\hspace*{0.5mm} $^h$ Bolometric amplitude (mag).
\hspace*{0.5mm} $^i$ Amplitude of logarithmic static gravity.  
\hspace*{0.5mm} $^j$ Amplitude of logarithmic effective gravity.
\hspace*{0.5mm} $^k$ Surface temperature variation (K).   
\hspace*{0.5mm} $^l$ Effective temperature variation (K).
\hspace*{0.5mm} $^m$ Total kinetic energy (erg).}
\end{deluxetable}


\begin{deluxetable}{crrcrrc}
\tablecaption{Analytical relations for the effective temperature of 
fundamental edges at fixed chemical composition.  
$\log T = \alpha + \beta \log L/L_\odot$ \label{tbl-5}}
\tablewidth{0pt}
\tablehead{
\colhead{Z\tablenotemark{a}}&
\colhead{$\alpha$\tablenotemark{b}}&
\colhead{$\beta$\tablenotemark{c}}&
\colhead{$\sigma$\tablenotemark{d}}&
\colhead{$\alpha$\tablenotemark{b}}&
\colhead{$\beta$\tablenotemark{c}}&
\colhead{$\sigma$\tablenotemark{d}} \nl 
        & \multicolumn{3}{c}{FBE\tablenotemark{e}} & \multicolumn{3}{c}{FRE\tablenotemark{f}}   
}
\startdata
0.004 & 3.905 & -0.035 & 0.004 & 3.995 & -0.080 & 0.008 \nl
      &$\pm$0.004\tablenotemark{g}&$\pm$0.004& &$\pm$0.008&$\pm$0.006 &\nl
0.008 & 3.919 & -0.043 & 0.008 & 4.030 & -0.094 & 0.008\nl
      &$\pm$0.008&$\pm$0.006&  &$\pm$0.009&$\pm$0.005 & \nl
0.02 & 3.959 & -0.061 & 0.006 & 4.125 & -0.124 & 0.011\nl
      &$\pm$0.006&$\pm$0.004& & $\pm$0.011&$\pm$0.009 &  \nl
\enddata
\tablenotetext{a}{Metal content.
\hspace*{0.5mm} $^b$ Zero point.
\hspace*{0.5mm} $^c$ Luminosity coefficient
\hspace*{0.5mm} $^d$ Standard deviation.
\hspace*{0.5mm} $^e$ Fundamental blue edge.
\hspace*{0.5mm} $^f$ Fundamental red edge.
\hspace*{0.5mm} $^g$ Errors on coefficients.} 
\end{deluxetable}


\begin{deluxetable}{crrrrc}
\tablecaption{Fundamental pulsation relations at fixed chemical composition.  
$\log P = \alpha + \beta \log T_e + \gamma \log M/M_\odot +
\delta \log L/L_\odot$ \label{tbl-6}}
\tablewidth{0pt}
\tablehead{
\colhead{Z\tablenotemark{a}} &
\colhead{$\alpha$\tablenotemark{b}} &
\colhead{$\beta$\tablenotemark{c}} &
\colhead{$\gamma$\tablenotemark{d}}&
\colhead{$\delta$\tablenotemark{e}}&
\colhead{$\sigma$\tablenotemark{f}}
}
\startdata
0.004 & 10.971 & -3.387  & -0.813 & 0.929 & 0.008  \nl
      &$\pm$0.008\tablenotemark{g}&$\pm$0.034& $\pm$0.030 & $\pm$0.008 &\nl
0.008 & 10.557 & -3.279 & -0.795 & 0.931 & 0.008   \nl
      &$\pm$0.008 &$\pm$0.036& $\pm$0.032 & $\pm$0.008 &\nl
0.02 & 9.874 & -3.108 & -0.767 & 0.942 & 0.008  \nl
      &$\pm$0.008&$\pm$0.042& $\pm$0.039 & $\pm$0.010 & \nl
\enddata
\tablenotetext{a}{Metal content.
\hspace*{0.5mm} $^b$ Zero point.
\hspace*{0.5mm} $^c$ Effective temperature coefficient.
\hspace*{0.5mm} $^d$ Mass coefficient.
\hspace*{0.5mm} $^e$ Luminosity coefficient.
\hspace*{0.5mm} $^f$ Standard deviation.  
\hspace*{0.5mm} $^g$ Errors on coefficients.}
\end{deluxetable}